\documentclass[lettersize,journal]{IEEEtran}
\usepackage{amsmath,amsfonts}
\usepackage{amsmath}
\usepackage{amssymb}
\usepackage{algorithmic}
\usepackage{algorithm}
\usepackage{array}
\usepackage[caption=false,font=normalsize,labelfont=sf,textfont=sf]{subfig}
\usepackage{textcomp}
\usepackage{stfloats}
\usepackage{url}
\usepackage{verbatim}
\usepackage{graphicx}
\usepackage{cite}
\usepackage{hyperref}
\usepackage{xcolor}
\definecolor{darkgreen}{RGB}{0,0,0}
\begin{document}

\title{Power Flow Solvability with Volt-Var Controlled Inverter-Based Resources}
\author{Taha Saeed Khan,~\IEEEmembership{Student Member,~IEEE,}
        Hamidreza Nazaripouya,~\IEEEmembership{Senior Member,~IEEE}
\thanks{Taha Saeed Khan and Hamidreza Nazaripouya are with the School of Electrical Computer Engineering, Oklahoma State University, Stillwater 74075, OK, USA}
\thanks}

\markboth {PREPRINT}
{Shell \MakeLowercase{\textit{et al.}}: A Sample Article Using IEEEtran.cls for IEEE Journals}

\maketitle
\begin{abstract}
This paper establishes a sufficient condition for guaranteeing power flow solvability in distribution grids with inverter-based resources (IBRs) operating under IEEE 1547-compliant Volt-Var control. While designed to improve voltage profiles, reactive power injection can drive the system toward its operational limits. Under these stressed conditions, any further incremental reactive power injection can trigger voltage collapse—the point at which a power flow solution ceases to exist. In this paper, by leveraging a phasor-based voltage representation, the power flow equations with Volt-Var control are developed in the complex fixed point form, enabling a compact formulation and the rigorous application of fixed-point theorems. Addressing the challenges posed by the non-holomorphicity of the complex power flow equations due to the Volt-Var function's dependence on voltage magnitude, the solvability conditions are then developed using the Brouwer fixed-point theorem. The proposed conditions are validated through simulations on distribution test feeders, with a primary focus on their application to real-time decision-making for voltage regulation services.

\end{abstract}
\begin{IEEEkeywords}
Volt-Var control, Solvability Analysis, Distribution Grids, IBRs, Voltage Regulation, Fixed Point Method.
\end{IEEEkeywords}

\section*{Nomenclature}
{\small 
\begin{IEEEdescription}[\IEEEusemathlabelsep\IEEEsetlabelwidth{$V_1,V_2,V_5,V_2,V_5,V_2,V_2$}]
\item[$v = (v_1, v_2, \ldots, v_N)^T$] $v_k$ is the positive-sequence complex voltage at bus \textit{k}.
\item[$s = (s_1, s_2, \ldots, s_N)^T$] $s_k$ is the complex nodal power injected into bus \textit{k}.
\item[$m = (jm_1, \ldots, jm_N)^T$] $m_k$ is the slope of the voltage magnitude dependent reactive power exchanged at bus \textit{k}.
\item[Bus 0] Slack bus.
\item[$v_0, s_0$] Slack bus positive-sequence complex voltage and power. 
\item[$Y$] Positive-sequence nodal admittance matrix.
\item[$Y_{LL}$] Square submatrix of $Y$, excluding the slack bus.
\item[$w = (w_1, w_2, \ldots, w_N)^T$] $w_k$ is the positive-sequence no-load complex voltage at bus \textit{k}.
\item[$u= W^{-1}v$,  $W = \text{diag}(w)$] Normalized positive-sequence voltages.
\item[$\bar{z}$] The complex conjugate of the complex number $z$. 
\item[$\|A\|_{\infty} \triangleq \max_k \sum_l |A_{kl}|$] The \( \ell_{\infty} \) norm of $A$. 
\item[$\odot$] Element-wise (Hadamard) product operator; 
for matrices $A$ and $B$ of equal dimension,
$(A \odot B)_{ij} = a_{ij} b_{ij}$. 
\end{IEEEdescription}
}

\section{Introduction}
\IEEEPARstart{P}{ower} distribution grids are conventionally engineered to supply reliable power to consumers.
Exceeding the capacity of a power distribution system by integrating additional load or generation can lead to voltage collapse \cite{SimpsonPorco2016}. Guaranteeing the existence of a power flow solution \cite{Overbye1994} for a given load and generation indicates that the distribution grid is capable of serving loads without causing voltage collapse. 
When a power flow solution ceases to exist, it indicates that the system has reached its maximum transferable power limit. This event is a saddle-node bifurcation, the point at which the operating point ceases to exist, hence leading to voltage collapse \cite{Kwatny1986} \cite{Weng2022} \cite{Wang2017}.

Distribution grids exhibit diverse characteristics, influenced by geographical, demographic, and climatic factors. These differences are reflected in the electrical load patterns of consumers and the impedance properties of the distribution system. Given the diverse operating conditions in a distribution grid, it is crucial for power system operators to understand the allowable operating ranges and security regions necessary to ensure secure and feasible grid operation.

With the increasing complexity of load and generation patterns, distribution grids face new challenges: the most important is the need for effective voltage regulation \cite{Review2019}, a task complicated by uncertain load patterns, intermittent renewable generation, and the widespread inclusion of fast electric vehicle (EV) chargers.

One method proposed by IEEE 1547 \cite{IEEE15472018} for regulating voltages using IBRs includes injecting reactive power. A widely adopted approach for this is the use of Volt-Var functionality, where each inverter in the grid is programmed to inject or absorb reactive power as a function of its nodal voltage magnitude. These functions are configured by setting slope intercept points in inverter settings \cite{{Narang2021}}. Programming inverters with Volt-Var curves leads to voltage-dependent reactive power generation at various nodes throughout the grid. However, this Volt-Var-based reactive power injection can cause several issues, including control instability, voltage oscillations, and lack of adaptability to changing grid conditions, as highlighted in \cite{Review2019}. 

\begin{figure}[htbp]
\centering
\includegraphics[width=0.3\textwidth]{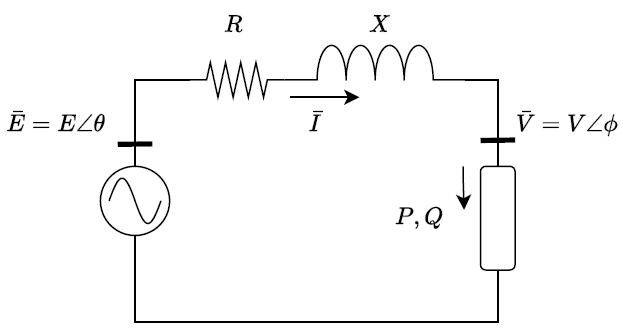}
\caption{A single load supplied by an infinite bus via a power line.}
\label{fig:simplelosslesscircuit}
\end{figure}

With increasing levels of active power exchanged with the grid, reactive power support can serve two purposes: It allows the grid to feed more load, commonly depicted by an extended nose curve \cite{VanCutsem1998}. Secondly, it helps to bring the voltage profile of the grid within an acceptable range. However, similar to the limits on active power loading in distribution systems, there is also an upper bound on the amount of reactive power that can be injected. The upper limits for the exchange of active power $P$ and reactive power $Q$ in a simple two-bus system, shown in Fig.~\ref{fig:simplelosslesscircuit}, are given by (\ref{eq:eins}), where $E$ is the ideal voltage source representing the infinite bus. $R$ and $X$ represent the series resistance and reactance, respectively\cite{VanCutsem1998}.

\begin{equation}
    \begin{aligned}
P_{\text{max}}(Q=0) = \frac{E^2}{2X^2}\left(-R + \sqrt{R^2+X^2}\right) \\
Q_{\text{max}}(P=0) = \frac{E^2}{2R^2}\left(-X + \sqrt{R^2+X^2}\right)
    \end{aligned}
    \label{eq:eins}
\end{equation}

Exceeding these limits can trigger voltage collapse. Hence, when making operational decisions, each scenario must be validated to ensure the existence of a feasible power flow solution.

With the rapid deployment of inverter-based resources across the distribution grid, Volt-Var functionality can be effectively used to maintain the voltage profile across the grid, particularly when high-demand loads are connected, such as EV chargers. However, with Volt-Var curve-based reactive power support, the total reactive power injection across the system cannot be predetermined, as the reactive power supplied at each node is adjusted based on its local voltage conditions. Therefore, careful tuning of these curves is essential to prevent excessive reactive power injection, especially during high grid loading. Failure to do so may push the system toward voltage collapse.

Optimizing the Volt-Var control curve requires running power flow analyses across multiple scenarios to ensure reliable and effective grid operation. Guaranteeing the existence of a valid power flow solution around the operating point can eliminate the need to repeatedly solve complex power flow equations. This approach significantly improves computational efficiency, enabling faster decision-making in real-time applications. These challenges motivate the need for analytical tools that can certify the existence of a power flow solution under voltage-dependent control actions, without relying on repeated numerical power flow computations.

The remainder of the paper is organized as follows. Section II reviews existing relevant approaches and discusses the analytical advantages of conducting analysis in the complex domain, which serves as a foundation to introduce the non-holomorphic characteristics of voltage dependent power exchanges and their impact on analyzing solvability with Volt-Var control. The paper proceeds in Section III with the formulation of the power flow problem considering Volt-Var control. Section IV provides the solvability certificate for systems with voltage-dependent power exchanges based on current or past data of operating conditions. In Section V, the proposed solvability certificate is tested on IEEE test case scenarios, with the conclusion presented in Section VI.

\section{Solvability Challenges in Power Flow}

\textcolor{darkgreen}{
The solvability of power flow in distribution grids has long been a central focus of research, with particular emphasis on assessing how close a given operating point is to the system’s maximum loading capacity. Traditional approaches address this question by explicitly solving the power flow equations and placing significant effort on improving the computational speed of these solvers.
\newline 
\hspace*{\parindent} Among these methods, continuation power flow (CPF) \cite{Ajjarapu1992, Nirbhavane2021} remains the standard technique for accurately tracing the maximum loading point. However, CPF is computationally intensive due to its predictor–corrector structure, which requires repeatedly solving the nonlinear power flow equations at each incremental loading step. As a result, the existence of a solution is verified directly through numerical convergence at every iteration, leading to a substantial computational burden that can limit its applicability in real-time or large-scale scenarios.
\newline 
\hspace*{\parindent} Motivated by this challenge, improving the computational efficiency of power flow analysis in inverter-dominated networks has attracted significant research attention. A Jacobian-based phasor-domain power flow solver that explicitly incorporates inverter-based resource models using a Modified Augmented Nodal Analysis (MANA) formulation has been proposed in \cite{zahid1}, demonstrating reductions in computation time while maintaining accurate short-circuit current calculations. A general fast power flow algorithm that extends the classical DC formulation to both transmission and distribution networks with high line resistance is proposed in \cite{R9_Wang}. Nevertheless, these approaches still fundamentally rely on explicitly solving the power flow equations. This motivates alternative frameworks that assess feasibility and solvability directly, without the need to iteratively solve the power flow problem.
}
\subsection{Shift of Focus from Real to Complex Domain}
Solvability analysis emerges as a vital tool in facilitating real-time decision making in active distribution grids\cite{Wang2017}. Early foundational studies \cite{DOE1979, ChiangBaran1990} formulated the power flow problem predominantly in the real domain to develop solvability conditions. The pioneering work on solvability certificates by \cite{HnyiliczaLee1975}, followed by the quantitative framework introduced in \cite{Galiana1983}, highlighted the quadratic nature of the power flow equations and delineated feasible regions for steady-state operation.

While these approaches provide valuable insights, they inherently treat voltage magnitude and angle separately, making it difficult to construct fixed-point mappings that preserve the analytical structure of the AC power flow equations \cite{simpsonP, Bolognani2016}. This difficulty arises primarily from the nonlinear trigonometric representation of voltage magnitudes and phase angles in power flow equations, which complicates the satisfaction of essential conditions required by fixed point theorems, such as compactness, self-mapping, and contraction, thus hindering the direct application of analytical fixed point criteria for establishing solvability \cite{FP_Theorems}.

Existing methods based on Banach’s theorem \cite{Bazrafshan2018, Wang2018} and Brouwer’s theorem \cite{Nguyen2019} have established sufficient solvability conditions for constant-power injections. However, these approaches do not extend to scenarios where power injections are voltage-magnitude-dependent, such as those governed by Volt-Var control functions. The Newton-Raphson method \cite{Haque1996} has been used to solve the power flow problem for a variety of voltage-magnitude-dependent models, but no solvability certificate has yet been provided for such loads.

\subsection {Challenges with Complex Formation}
The initial challenge in formulating the power flow problem as a complex fixed-point equation lies in the fact that the mapping is generally nonholomorphic. A holomorphic function is a complex function that is complex differentiable in a neighborhood of each point in its domain, implying that it is smooth and infinitely differentiable. Unlike real functions that can be differentiable at a point but not smooth everywhere, a complex function that is differentiable in a complex coordinate space is smooth (\( C^{\infty} \)) throughout its domain and has derivatives of all orders. The smoothness and ability to express a function as a convergent power series stem from satisfying the Cauchy-Riemann equations. This condition is stricter than real differentiability, as complex numbers inherently represent two real-valued variables, interpreted as the real and imaginary components. Hence, they are particularly well-suited for modeling complex power, voltage, and impedance in power systems.

To address the non-holomorphic nature of the power flow problem, the holomorphic embedding load flow method (HELM) \cite{Trias2012} was employed. This technique was particularly developed to overcome the shortcomings of conventional iterative numerical methods (e.g., Newton-Raphson, Gauss-Seidel), as their convergence cannot be guaranteed for a power system under critical loading, even when a solution exists. In particular, the holomorphic embedding method \cite{Trias2012} relies on the derivation of power series \cite{BakerGravesMorris1996} in a complex form, prompting the question of whether voltage-magnitude-dependent load and generation models, such as those governed by Volt-Var control, can be represented as power series, considering that the derivation of power series requires the underlying function to be differentiable.

The problem of differentiability in the power flow problem was also observed in \cite{Cui2019}, and a Schwarz lemma-based solution for the multivariable problem under constant power injections was proposed. The latest work in \cite{Weng2022} applied the Kantorovich fixed point theorem to guarantee the existence and uniqueness of the power flow solutions; however, the approach also relies on the differentiability of the underlying function.

The power flow equation, when formulated with constant loads, is non-holomorphic as it involves the complex conjugate of the voltage variable \( v \). Wirtinger Calculus \cite{Wang2017} can be employed to deal with differentiability when conjugate variables are the only source of non-holomorphicity. An approximate power series around the operating point can also be used to address this issue\cite{Wang2018}. Another technique, known as the holomorphic embedding method, was also proposed to handle the complex conjugate part \cite{Trias2012}. However, a key challenge in networks with inverters operating in Volt-Var control mode is that the power flow equations are functions of voltage magnitudes rather than complex voltage variables, introducing a unique non-holomorphic property.

This poses a challenge for solvability analysis, particularly when applying the fixed-point theorem that requires the contraction property to hold. For example, the Banach fixed point theorem requires demonstrating that the system's mapping is a contraction. Although differentiability is not a requirement, the standard methods for proving contraction and deriving practical solvability certificates rely on bounding the function's derivative. This presents a major hurdle, as the voltage-magnitude dependencies render the power flow equations non-differentiable (non-holomorphic), making these conventional techniques inapplicable.

\textbf{Remark 1:} The function \( f(z) = |z| \), where \( z \) is a complex variable, is continuous everywhere, but is not differentiable anywhere. Therefore, an approximate power series representation of the power flow equations involving this function in the form of complex variables cannot be developed, except at the zero voltage vector. Thus, it can be concluded that for voltage-dependent power sources, proving contraction is a challenging task as the corresponding functions are not differentiable.

\textcolor{darkgreen}{
\subsection {Contribution}
In this work, Brouwer’s fixed-point theorem is leveraged to develop a solvability certificate for power flow equations that incorporate voltage-dependent reactive power injections.
Existing research on the solvability of power flow predominantly concentrates on ZIP (constant-power, constant-current, and constant-impedance) loads \cite{Bernstein2015,Nguyen2019,Cui2019,Chen1991,Wang2018}\cite{Bazrafshan2018}. This paper derives a generalized solvability certificate that simultaneously accommodates both Volt-Var characteristics and constant power load models. This enables the prescreening of operating conditions for which voltage regulation functions are guaranteed to converge, even in highly stressed or weakly supported grid conditions.
Hence, the primary contributions of this work are as follows.
\begin{itemize}
    \item It embeds voltage-magnitude-dependent power injections, characterized by Volt-Var control, into a tractable fixed-point framework.
    \item It establishes checkable conditions using the Brouwer fixed point theorem that guarantee the existence of a power flow solution under voltage-magnitude-dependent power injections characterized by Volt-Var control.
\end{itemize}
}

\section{Mathematical Modeling}
\subsection{Volt-Var Function}
\begin{figure}[htbp]
\centering
\includegraphics[width=0.4\textwidth]{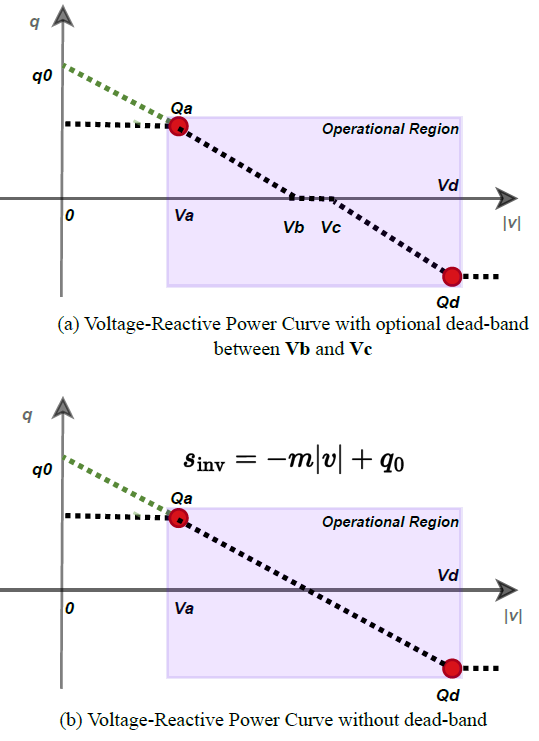}
\caption{Voltage-magnitude-dependent reactive power generation and absorption curves. The deadband is optional in the inverter setting and can be removed by setting \( V_b = V_c \).
}

\label{fig:voltvargen}
\end{figure}

In this section, the Volt-Var function and its manifestation in the fixed-point-based power flow equation are provided. Volt-Var curves are characterized by a slope, saturation points, and an optional dead band, shown in Fig.~\ref{fig:voltvargen}(a) \cite{Narang2021}. When an inverter is operating in Volt-Var mode, its reactive power response can be approximated as a linear function of the voltage magnitude in the vicinity of the operating point. Hence, an inverter operating in the Volt-Var mode can be effectively modeled using the slope-intercept form, as shown in Fig.~\ref{fig:voltvargen}(b).

The linear Volt-Var based power injection around the reference point at bus \textit{k} can hence be formulated in a complex form as (\ref{eq:one}):
\begin{equation}
    s_{\text{inv},k} =j(- m_{k} |v_{k}| + q_{0,k}) \triangleq  Q_{\text{Volt-Var}}
    \label{eq:one}
\end{equation}
where $j$ is the symbol used to denote the imaginary unit, $s_{\text{inv},k}$ represents the reactive power injected by an inverter at bus $k$. $m_{k}$ and $q_{0,k}$ represent the slope and \textit{q} axis intercept, respectively, resulting from programming the inverter at $\textit{k}^{\text{th}}$ bus with Volt-Var functions. $|v_{k}|$ is voltage magnitude at $k^{\text{th}}$ bus.

In addition to power injection by inverters, consider a constant load/generation $s_{k}^{PQ}$ connected at the $k^{\text{th}}$ bus with $p_{k}$ and $q_{k}$ being the net constant active and reactive power components, respectively, as shown in (\ref{eq:two}):
\begin{equation}
    s_{k}^{PQ} = p_{k}+ jq_{k}
    \label{eq:two}
\end{equation}

Hence, the net power injection $s_{k}$ at bus \textit{k} is expressed as (\ref{eq:3}) and (\ref{eq:4}):
\begin{equation}
    s_{k} = s_{\text{inv},k} + s_{k}^{PQ}
    \label{eq:3}
\end{equation}
\begin{equation}
    s_{k} = p_{k} + j(- m_{k} |v_{k}| + q_{0,k} + q_{k})
    \label{eq:4}
\end{equation}

Separating the voltage dependent and independent components in the complex power results in (\ref{eq:5}):
\begin{flalign}
    s_{k} &= - (jm_{k} ) |v_{k}| + (p_{k} + jq_{0,k} + jq_{k}) \nonumber \\
    &= - (jm_{k} ) |v_{k}| + s_{k,\text{const}}
    \label{eq:5}
\end{flalign}
where the voltage independent component is defined as:
\begin{equation}
s_{k,\text{const}}\triangleq p_{k} + jq_{0,k} + jq_{k}
\label{eq:def}
\end{equation}

\subsection{Power Flow Equation}
Consider a distribution network modeled using its positive-sequence equivalent, comprising $N$ PQ buses and a single slack bus with a complex voltage fixed at 1 p.u. Under constant power loading/generation $s^{PQ}\triangleq (s_{1}^{PQ}, s_{2}^{PQ}, \ldots, s_{N}^{PQ})^T$, the fixed-point formulation of the power flow equations can be expressed as (\ref{eq:FixPF}) \cite{Wang2018}:
\begin{equation}
v = w + Y_{LL}^{-1} \text{diag}(\bar{v})^{-1} \overline{ s^{PQ}} \triangleq G(v)
\label{eq:FixPF}
\end{equation}

Here, $Y_{LL}$ is the reduced admittance sub-matrix obtained by removing the slack bus from the full admittance matrix $Y$. Following the \textit{implicit} $Z_{\text{bus}}$ \textit{formulation} presented in \cite{Chen1991}. The admittance matrix $Y$ can be partitioned as:
\begin{equation}
Y = 
\begin{bmatrix}
Y_{00} & Y_{0L} \\
Y_{L0} & Y_{LL}
\end{bmatrix},
\label{eq:Y_partition}
\end{equation}
where $Y_{00}$ is a scalar, $Y_{0L}$ is a $1 \times N$ row vector, $Y_{L0}$ is an $N \times 1$ column vector, and $Y_{LL}$ is an $N \times N$ matrix.

The vector \( w \), defined  as the zero-load voltage of the grid, is given by (\ref{eq:w}):
\begin{equation}
w \triangleq -Y_{LL}^{-1}Y_{L0}
\label{eq:w}
\end{equation}

By defining $s_{\text{const}} = (s_{1,\text{const}}, \ldots, s_{N,\text{const}})^T$, the power flow equation incorporating Volt-Var functionality and constant power injections, as defined in (\ref{eq:5}), can be rewritten as (\ref{eq:PF}): 
\begin{equation}
v = w + Y_{LL}^{-1} \text{diag}(\bar{v})^{-1} (\overline{-\text{diag}(m) |v| + s_{\text{const}}}) \triangleq F(v)
\label{eq:PF}
\end{equation}
To solve the power flow equation in (\ref{eq:PF}), it can be reformulated into the iterative structure presented in (\ref{eq:iterPF}):
\begin{equation}
v^{(l+1)} = w + Y_{LL}^{-1} \text{diag}(\bar{v}^{(l)})^{-1} (\overline{-\text{diag}(m) |v^{(l)}| + s_{\text{const}}})
\label{eq:iterPF}
\end{equation}

\subsection{Normalized Form}

To facilitate further analysis, the power flow equation (\ref{eq:PF}) is normalized. Let \( W \triangleq \text{diag}(w) \) and \( u \triangleq W^{-1}v \), where \( u \) is the normalized voltage vector. Substituting \( v = Wu \) into (\ref{eq:PF}) yields:

\begin{align}
Wu = w + Y_{LL}^{-1}\overline{W}^{-1} \text{diag}(\overline{u})^{-1} (\overline{- \text{diag} (m)|Wu| + s_{\text{const}}})
\label{eq:13o}
\end{align}

\textcolor{darkgreen}{
Defining, $ W^{*}\triangleq W^{-1}Y_{LL}^{-1}\overline{W}^{-1} $
\begin{align}
u =  W^{-1}w +  W^{*} \text{diag}(\overline{u})^{-1} (\overline{- \text{diag} (m)|W||u| + s_{\text{const}}})
\label{eq:13a}
\end{align}
Defining,
\begin{align*}
\mathcal{M} & \triangleq m \odot |w|
\\
M & \triangleq \text{diag} (m \odot |w|)
\end{align*}
The equation (\ref{eq:13a}) can be written as (\ref{eq:13b}):
}
\begin{equation}
u = \mathbf{1} + W^{*}\text{diag}(\overline{u})^{-1} (\overline{- M|u| + s_{\text{const}}}) \triangleq \Tilde{F}(u)
\label{eq:13b}
\end{equation}
where \( \mathbf{1} = (1, 1, \ldots, 1)^T \) is the unity vector.
Now the following iteration of $\Tilde{F}(u)$ can be written as: 
\begin{equation}
u^{(l+1)} = \mathbf{1} + W^{*} \text{diag}(\overline{u^{(l)}})^{-1}(\overline{-M|u^{(l)}| + s_{\text{const}}})
\label{eq:NormPF_iter}
\end{equation}

\section{ Power Flow Solvability with Volt-Var Function}

This section provides the solvability conditions and their proof based on the Brouwer fixed point theorem\cite{BrouwerReference}. Brouwer's Fixed Point Theorem has an advantage, especially for analyzing complex power flow equations, as it only demands the continuity of the function, without necessitating differentiability. Although the theorem does not guide on how to locate the solution point(s), it ensures the existence of at least one fixed point. The theorem offers a promising way to verify the problem solvability, it does not provide any information on the potential convergence or divergence of the iterative methods that may be used to find a solution.

\textbf{Theorem 1 (Brouwer's Fixed Point Theorem)}\textbf{:} In a finite-dimensional, compact, convex subset of a Euclidean space, any continuous function in \( \mathbb{C}^n \) that maps the set to itself has at least one fixed point.

Thus, the following three conditions need to be satisfied for the application of Brouwer's Fixed Point theorem on the iterative form (\ref{eq:NormPF_iter}):
\begin{enumerate}
    \item Euclidean space: Since the problem under consideration (i.e., the power flow equation) is a finite-dimensional problem, the space under consideration is a Euclidean space.
    \item Continuity: Although the power flow equation with the Volt-Var function is not differentiable, it is continuous.
    \item Self-mapping on compact and convex set: This is the only condition that needs to be proved. Thus, the goal is to find a compact convex set within which the iteration (\ref{eq:NormPF_iter}) maps the set to itself.
\end{enumerate}

\subsection{Solvability Certificate}

\textcolor{darkgreen}{
\textbf{Theorem 2:}  
Let \( \hat{v} \) be a solution to the power flow problem (\ref{eq:PF}) for complex power injection \(\hat{s}\triangleq - \text{diag} (\hat{m}) |\hat{v}| + \hat{s}_{\text{const}}\). Consider another power injection candidate given by \(s\triangleq - \text{diag} (m) |v| + s_{\text{const}}\). To evaluate the solvability of the candidate power injection $s$, we introduce the operator
\(\xi(\hat{s}) \triangleq \left\| W^{*}\text{diag}(\overline{\hat{s}}) \right\|_{\infty}\), \(\xi(\mathcal{M}) \triangleq \left\| W^{*}\,\text{diag}(\overline{\mathcal{M}}) \right\|_{\infty}\), and \(\xi(s_{const}-\hat{s} - M|\hat{u}|) \triangleq \left\| W^{*}\text{diag}(\overline{s_{const}-\hat{s} - M|\hat{u}|}) \right\|_{\infty}\), to capture the maximum weighted influence of known power injections, Volt-Var slopes, and the residual power injection mismatch on the normalized voltage update $\Tilde{F}(u)$, respectively. Furthermore, let \(u_{\text{min}} \triangleq \min_j \left|\frac{\hat{v}_j}{w_j}\right|\) be the minimum normalized voltage magnitude associated with the known feasible operating point \( \hat{v} \).
If the candidate power injection satisfies the following sufficient conditions:
}

\begin{equation}
\left ( u_{min} - \frac {\xi(\hat{s})}{u_{min}} - {\xi(\mathcal{M})}\right) > 0
\label{eq:cond_a}
\end{equation}

\begin{equation}
\Delta \triangleq \left ( u_{min} - \frac {\xi(\hat{s})}{u_{min}} - {\xi(\mathcal{M})}\right)^{2} -4\xi(s_{\text{const}} - \hat{s} - M|\hat{u}|) > 0
\label{eq:cond_b}
\end{equation}

Then there exists a solution \( v \in \mathcal{D} \) to the power flow problem (\ref{eq:PF}), where \( \mathcal{D} \triangleq \{v : |v_j - \hat{v}_j| \leq \rho |w_j|, j = 1, \ldots, N\} \) with
\( \rho \triangleq \frac{\left ( u_{min} - \frac {\xi(\hat{s})}{u_{min}} - {\xi(\mathcal{M})}\right) - \sqrt{\Delta}}{2}
\)

The following lemma is a direct consequence of the above theorem:

\textbf{Lemma}: Suppose that the pair \((\hat{v},\hat{s})\) and power injection candidate given by \(s\triangleq - \text{diag} (m) |v| + s_{\text{const}}\) satisfies (\ref{eq:cond_a}) and (\ref{eq:cond_b}). Then \( \Tilde{F}\) is a self-mapping of \( u \triangleq W^{-1}v \) on $\Tilde{\mathcal{D}}$, where 

\begin{equation}
\Tilde{\mathcal{D}} \triangleq \{u : |u_j - \hat{u}_j| \leq \rho, j = 1, \ldots, N\} \label{SMDomain}
\end{equation}
and  \(\hat{u}_j = \frac{\hat{v}_j}{w_j}\).

{\small 
\textbf{Proof:}
Now the focus is on proving self-mapping of equation (\ref{eq:NormPF_iter}) on the compact convex set $\Tilde{\mathcal{D}}$.
Since \((\hat{v},\hat{s})\) satisfies the equation (\ref{eq:PF}), from (\ref{eq:13b}) we have:
\begin{align}
\hat{u} = \mathbf{1} + W^{*}\text{diag}(\overline{\hat{u}})^{-1} (\overline{\hat{s}})
\label{eq:second}
\end{align}

For any new power injection vector \(s = - M|u| + s_{const}\), the power flow equation is given by (\ref{eq:13b}). Hence, the difference between (\ref{eq:13b}) and (\ref{eq:second}) can be written as:

\begin{align*}
\begin{split}
&\Tilde{F}(u) - \hat{u} =\\
&W^{*}\bigg( \text{diag}(\bar{u})^{-1}\left(\overline{- M|u| + s_{const}}\right)- \text{diag}(\bar{\hat{u}})^{-1}\left(\overline{\hat{s}}\right)\bigg)
\end{split}
\end{align*}

The objective is to demonstrate that there exists a radius \( r \) such that if \( |{u}_i - \hat{u}_i| \leq r \) then \( |\Tilde{F}(u)_i - \hat{u}_i| \leq r \) for all \( i \).
\begin{align*}
|\Tilde{F}(u)_i - \hat{u}_i| &= \left| \sum_{j} \left( W^{*} \right)_{ij} \left( \frac{(\overline{- M|u| + s_{const}})_j}{\Bar{u}_j} - \frac{\overline{\hat{s}_{j}}}{\Bar{\hat{u}}_j} \right) \right|
\end{align*}

\begin{align*}
 &= \left| \sum_{j} {W^{*}}_{ij} \left( \frac{(\overline{- M|u|)_{j}\hat{u}}_{j} + \overline{(s_{const})_{j}\hat{u}_{j}} - \overline{\hat{s}_{j}u_j}}{\overline{{u_j}\hat{u}_j}} \right) \right|
\end{align*}

Adding and subtracting: \(\frac{\overline{\hat{s}_{j}{\hat{u}_j}}}{\overline{u_j\hat{u}_j}} \):
\textcolor{darkgreen}{
\begin{align*}
\begin{split}
&= \left| \sum_{j} {W^{*}}_{ij} \left( \frac{\overline{(- M|u|)_{j}\hat{u}_{j}} + \overline{\hat{u}_{j}(s_{j,const}-\hat{s}_{j})}}{\overline{u_j\hat{u}_j}} \right. \right. +\\
&\qquad \qquad \qquad \left. \left.  \frac{\overline{\hat{s}_{j}(\hat{u}_{j}-u_j)}}{\overline{u_j\hat{u}_j}} \right) \right|
\end{split}
\end{align*}
}

Adding and subtracting \(\frac{\overline{(M|\hat{u}|)}_{j}}{\overline{u}_j} \):
\textcolor{darkgreen}{
\begin{align*}
\begin{split}
&= \left| \sum_{j} {W^{*}}_{ij} \left( \frac{-\overline{{(M|u|)}}_{j} + \overline{(s_{j,const}-\hat{s}_{j})}}{\overline{u_j}} \right. \right. + \\
&\qquad \qquad \qquad \left. \left. \frac{\overline{(M|\hat{u}|)}_{j} -\overline{(M|\hat{u}|)}_{j}}{\overline{u_j}}  + \frac{\overline{\hat{s}_{j}(\hat{u}_{j}-u_j)}}{\overline{u_j\hat{u}_j}} \right) \right|
\end{split}
\end{align*}
}

\textcolor{darkgreen}{
\begin{align*}
\begin{split}
&= \left| \sum_{j} {W^{*}}_{ij} \left( \frac{\overline{M_{jj}}(\overline{|\hat{u}_{j}|- |u_{j}|}) + \overline{(s_{j,const}-\hat{s}_{j} - M_{jj}|\hat{u}_{j}|})}{\overline{u_j}} \right. \right. + \\
&\qquad \qquad \qquad \left. \left.  \frac{\overline{\hat{s}_{j}(\hat{u}_{j}-u_j)}}{\overline{u_j\hat{u}_j}} \right) \right|
\end{split}
\end{align*}
}

\textcolor{darkgreen}{
Applying the triangle inequality upper-bounds the absolute value of the entire sum by the sum of the absolute values of its individual components:
}

\textcolor{darkgreen}{
\begin{align}
\begin{split}
& \leq \left\{ \sum_{j}\left| {W^{*}}_{ij} \left( \frac{\overline{M_{jj}}(\overline{|\hat{u}_{j}|- |u_{j}|}) + \overline{(s_{j,const}-\hat{s}_{j} - M_{jj}|\hat{u}_{j}|})}{\overline{u_j}} \right. \right. \right. + \\
& \left. \qquad \qquad \qquad \left. \left.  \frac{\overline{\hat{s}_{j}(\hat{u}_{j}-u_j)}}{\overline{u_j\hat{u}_j}} \right) \right| \right\}
\end{split}
\end{align}
}

\begin{align}
\leq \Biggl\{ \sum_{j}&\left| {W^{*}}_{ij} \left( \frac{\overline{M_{jj}}(\overline{|\hat{u}_{j}|- |u_{j}|})}{\overline{u_j}} + \frac{\overline{\hat{s}_{j}(\hat{u}_{j}-u_j)}}{\overline{u_j\hat{u}_j}} \right)\right|  \nonumber +\\
& \sum_{j}\left| {W^{*}}_{ij}\left(\frac{\overline{(s_{j,const}-\hat{s}_{j} - M_{jj}|\hat{u}_{j}|)}}{\overline{u_j}}\right) \right|
\Biggr\}
\label{slast_step}
\end{align}

\begin{align}
\le \Biggl\{
\sum_{j} & \left| {W^{*}}_{ij} \left( \frac{\overline{M_{jj}}(|\hat{u}_{j}|- |u_{j}|)}{\overline{u_j}}\right)\right| +  \nonumber \\
& \sum_{j}\left| {W^{*}}_{ij} \left(\frac{\overline{\hat{s}_{j}(\hat{u}_{j}-u_j)}}{\overline{u_j\hat{u}_j}} \right)\right| \nonumber +\\
& \sum_{j}\left| {W^{*}}_{ij}\left(\frac{\overline{(s_{j,const}-\hat{s}_{j} - M_{jj}|\hat{u}_{j}|)}}{\overline{u_j}}\right) \right|
\Biggr\}
\label{klast_step}
\end{align}

Now choosing \( r \) to be strictly less than \(u_{min}\) i.e. \( r < u_{min}\),  and  assuming all \( u_i \) are initialized within the radius \( r \) of \( \hat{u}_i\) i.e. they satisfy \( |u_i - \hat{u}_i| \leq r \),  then:
\begin{enumerate}
\item By definition of \(u_{min}\), \( |\hat{u}_j| \geq u_{min}\)
\item \( |u_j| \geq  u_{min} -r\) or \( |\overline{u}_j| \geq  u_{min} -r\)
\item By reverse triangular identity: \( r \geq |u_j - \hat{u}_j| \geq \left|      |u_j| - |\hat{u}_j| \right| \)
\end{enumerate}

}

By considering each term in (\ref{klast_step}) separately and using \(\xi(\hat{s})\), \(\xi(\mathcal{M})\), \(\xi(s_{const}-\hat{s} - M|\hat{u}|)\) and $u_{\text{min}}$ to upper bound them, one can obtain:

\begin{flalign}
&\sum_{j}\left| {W^{*}}_{ij} \left( \frac{\overline{M_{jj}}(|\hat{u}_{j}|- |u_{j}|)}{\overline{u_j}}\right)\right| \nonumber  \\
&\leq \sum_{j}\left| {W^{*}}_{ij} \overline{M_{jj}}\right| \frac {r}{(u_{min} -r)}
\leq \frac {\xi(\mathcal{M})  r}{(u_{min} -r)} \label{aa}
\end{flalign}
similarly,
\begin{flalign}
&\sum_{j}\left| {W^{*}}_{ij} \frac{\overline{\hat{s}_{j}(\hat{u}_j - u_j)}}{\Bar{u}_j{\Bar{\hat{u}}_j}} \vphantom{\sum_{j}} \right|
\leq  \sum_{j}\left| {W^{*}}_{ij} \overline{\hat{s}_{j}} \vphantom{\sum_{j}} \right|\frac {r}{(u_{min} -r)} \nonumber \\
&\leq \xi(\hat{s}) \frac {r}{(u_{min} -r)u_{min}} \label{bb}
\end{flalign}
and,
\begin{flalign}
&\sum_{j}\left| {W^{*}}_{ij}\left(\frac{\overline{(s_{j,const}-\hat{s}_{j} - M_{jj}|\hat{u}_{j}|)}}{\overline{u_j}}\right) \right| \nonumber  \\
& \leq  \sum_{j} \left|{W^{*}}_{ij}(\overline{(s_{j,const}-\hat{s}_{j} - M_{jj}|\hat{u}_{j}|)})\right|  \frac {1}{(u_{min} -r)} \nonumber \\
&\leq \xi((s_{const}-\hat{s} - M|\hat{u}|)) \frac {1}{(u_{min} -r)} \label{cc}
\end{flalign}

By combining (\ref{aa}), (\ref{bb}), and (\ref{cc}), the expression (\ref{slast_step}) can be written as (\ref{tlast_step}):
\begin{flalign}
&|\Tilde{F}(u)_i - \hat{u}_i|  
\leq \left\{ \frac {\xi(\mathcal{M})  r}{(u_{min} -r)}
+ \right. \nonumber \\
& \left. \xi((s_{const}-\hat{s} - M|\hat{u}|))  \frac {1}{(u_{min} -r)} + \xi(\hat{s}) \frac {r}{(u_{min} -r)u_{min}} \right\}
\label{tlast_step}
\end{flalign}

To fulfill the criteria of Self Mapping, the following condition must be met:
\begin{flalign}
\frac {\xi(\mathcal{M})  r}{(u_{min} -r)}
+  \frac {\xi(s_{const}-\hat{s} - M|\hat{u}|)}{(u_{min} -r)}+  \frac {\xi(\hat{s})r}{(u_{min} -r)u_{min}} \leq r
\end{flalign} 

\begin{flalign}
 \xi(\mathcal{M}) r + {\xi(s_{const}-\hat{s} - M|\hat{u}|)}+  \frac {\xi(\hat{s})r}{u_{min}}- r(u_{min} -r) \leq 0
\end{flalign} 

\begin{flalign}
 r^{2} - r \left ( u_{min} - \frac {\xi(\hat{s})}{u_{min}} - {\xi(\mathcal{M})}\right)  + \xi(s_{\text{const}} - \hat{s} - M|\hat{u}|)\leq 0 \label{dd}
\end{flalign}

Equation (\ref{dd}) is a quadratic equation in \textit{r}. Since \(\xi(s_{\text{const}} - \hat{s} - M|\hat{u}|) \) is positive, \(\left ( u_{min} - \frac {\xi(\hat{s})}{u_{min}} - {\xi(\mathcal{M})}\right)\) has to be positive for a positive \(r\) to exist.
Further, with the positive discriminant \( \Delta = \left ( u_{min} - \frac {\xi(\mathbf{\hat{s}})}{u_{min}} - {\xi(\mathcal{M})}\right)^{2} -4\xi(s_{\text{const}} - \hat{s} - M|\hat{u}|)\), existence of a positive \textit{r} is guaranteed. Hence, the following two conditions guarantee the existence of a solution: 
\begin{enumerate}
\item \(\left ( u_{min} - \frac {\xi(\hat{s})}{u_{min}} - {\xi(\mathcal{M})}\right) > 0\)
\item \(\left ( u_{min} - \frac {\xi(\hat{s})}{u_{min}} - {\xi(\mathcal{M})}\right)^{2} -4\xi(s_{\text{const}} - \hat{s} - M|\hat{u}|) > 0\)

\end{enumerate}

 Since a voltage solution in the near vicinity of the current operational point \((\hat{v},\hat{s})\) is of interest, the smaller solution is considered, which is provided by:
\( \rho = \frac {\left ( u_{min} - \frac {\xi(\hat{s})}{u_{min}} - {\xi(\mathcal{M})}\right) - \sqrt{\Delta}}{2}\)

\textbf{Remark 2}: Theorem 2 provides a sufficient condition for the existence of at least one fixed point of the function defined in the domain \( \mathcal{D}\), although it does not guarantee uniqueness. Furthermore, Brouwer’s theorem does not inherently guarantee that an iterative method will converge to any fixed point(s) within the domain \( \mathcal{D} \).

\textbf{Remark 3:}
The solvability certificate guarantees the existence of a solution to the nonlinear power-flow equations within a feasible domain~$\mathcal{D}$. Once the existence of a solution in $\mathcal{D}$ is guaranteed, system-level operational and control decisions can be made with assurance that a power flow solution exists in the admissible voltage region. If an explicit solution is required, then any standard power-flow solution method, including Newton-Raphson, HELM, or MANA, may be used to compute the exact solution within $\mathcal{D}$. Importantly, the proposed solvability certificate verifies only the existence of a solution in $\mathcal{D}$ and not the convergence or performance of any particular algorithm.

\textcolor{darkgreen}{
\subsection{Conservativeness and Solvability gaps}
Solvability gaps, defined as the distance between theoretical solvability bounds and practical insolvability points, are observed in numerical studies \cite{Porco} \cite{Bolognani2016}, which illustrates the conservative nature of the bounds used to ensure rigorous and broadly applicable solvability guarantees.
\newline
In this work, the resulting conservativeness is primarily attributed to the dependence on the minimum normalized voltage bound $u_{\min}$ and the induced norm quantities $\xi(\hat{s})$, $\xi(\mathcal{M})$, and $\xi((s_{const}-\hat{s} - M|\hat{u}|))$.
\newline
First, the dependence on the normalized minimum voltage bound $u_{\min}$ shapes the tightness of the solvability condition. It may impose additional restrictions on operating scenarios with heterogeneous voltage profiles due to localized low voltage conditions. Consequently, a gap may arise between the theoretical solvability bound and the actual solvability boundary observed in practice.
\newline 
Second, the quantities $\xi(\cdot)$ represent norm bounds that characterize worst-case weights stem from the network admittance structure (i.e., W*). Terms $\xi(\hat{s})$, $\xi(\mathcal{M})$,  and $\xi(s_{\text{const}}-\hat{s}-M|\hat{u}|)$ naturally depend on: (i) the selected operating point $\hat{s}$, (ii) the different magnitude of the Volt-Var slopes (i.e., m at each node) across the network, and (iii) the residual power-injection mismatch between the chosen operating point and the updated power injections, which comprise both incremental load changes and Volt–Var support, respectively. Variations in these factors can lead to noticeable changes in the corresponding $\xi$ values, thereby influencing the tightness of the resulting solvability bound.
}
\textcolor{darkgreen}{
\subsection{Computational Complexity and Scalability}
Conventional power flow solvers rely on iterative methods to solve the nonlinear power flow equations. When Volt-Var control is incorporated, the resulting voltage-dependent reactive power introduces additional complexity. In distribution systems with high inverter-based resource (IBR) penetration, this voltage-dependence can render the power flow equations ill-conditioned \cite{Milano}, often leading to slow convergence or even divergence of standard iterative solvers. 
In contrast, the proposed solvability certificate circumvents the need for iterative power flow solutions by verifying the existence of an equilibrium through a single, non-iterative evaluation.
\begin{enumerate}
\item  Computational Burden:
For radial networks, the nodal admittance matrix is highly sparse. By exploiting this sparsity and symmetry (e.g., using LU decomposition with complete Markowitz pivoting \cite{Markowitz}), the complexity of one power flow iteration is reduced to $\mathcal{O}(N)$.
\newline
In verifying the proposed solvabity conditions (\ref{eq:cond_a}) and (\ref{eq:cond_b}), computational complexity mainly stems from calculating $\xi(\hat{s})$, $\xi(\mathcal{M})$  and $\xi(s_{\text{const}} - \hat{s} - M|\hat{u}|)$. For the radial system, again, this complexity is inherently $\mathcal{O}(N)$ by exploiting sparsity and restricting the evaluation to leaf nodes only. Hence, the complexity of checking solvability conditions is equivalent to only one power flow iteration.
\item  Scalability and Robustness:
As the number of nodes $N$ increases, the computational time for evaluating the certificate scales linearly for radial systems (i.e.,  $\mathcal{O}(N)$)  as it primarily involves vector matrix products and norm calculations. Moreover, since the solvability condition relies on explicit, norm-based bounds rather than an iterative process, it avoids the slow convergence and divergence issues that typically burden power flow solvers.
\end{enumerate}
}
\textcolor{darkgreen}{
\subsection{Application}
The proposed conditions (\ref{eq:cond_a}) and (\ref{eq:cond_b}) have multiple applications. They can aid in determining how a distribution grid can be loaded incrementally with the Volt-Var function for voltage support. In such a scenario, these conditions assist in making real-time operational decisions on load increments and Volt–Var control setpoints. 
Once the existence of a solution in \(\mathcal{D} \) is guaranteed, system-level operational and control decisions can be made with confidence that a feasible equilibrium exists away from voltage-collapse regions, provided that \(\mathcal{D}\) lies within the feasible domain.
\newline 
\hspace*{\parindent} Due to load variations, multiple incremental loading scenarios are expected, Algorithm~1 checks the proposed solvability conditions across all anticipated scenarios and returns a flag of 1 if solvability is guaranteed for all scenario, and 0 otherwise. In a real-time Energy Management System (EMS) setting, a flag of 0 serves as an early-warning indicator, enabling operators to adjust Volt–Var control parameters—such as slope, deadband, or reactive power limits—in advance to restore solvability and prevent operation near voltage-collapse boundaries.
\newline 
\hspace*{\parindent} In the numerical illustrations presented in Section~V, a single incremental loading scenario is evaluated for each of the IEEE 37-bus and IEEE 123-bus test systems to demonstrate the screening process. These results highlight the real-time applicability of the proposed solvability conditions for active decision-making without repeatedly solving power-flow equations.
\newline 
\hspace*{\parindent} Further applications include determining how to effectively adjust the slopes of the Volt-Var function when the grid loading scenario varies over time. In addition, to improve computational efficiency in solving the optimal power flow problem involving Volt-Var controlled IBRs, the proposed conditions can be incorporated as constraints within a linearized power system model \cite{LinModel}.
}


\begin{algorithm}
\caption{\color{darkgreen}{Real-Time Solvability Screening for Different Loading Scenarios under Volt-Var Control}}
\label{alg:realtime_solvability_batch}
\color{darkgreen}{
\begin{algorithmic}[1]
\REQUIRE Known feasible operating point $(\hat{v}, \hat{s})$, network matrix $(W^*)$, an expected loading scenario set $\mathcal K = \{s_1, s_2,\dots, s_{K}\}$ with total $K$ scenarios, where each scenario consist of incremental load and Volt-Var set point $(M,s_{\mathrm{const}})$.
\ENSURE Solvability test for each scenario $s_k \in \mathcal K$, 
corresponding admissible radii $\{\rho^{(k)}\}$, and a solvability flag $\texttt{flag}$.
\vspace{0.2em}
\STATE Precompute scenario-invariant norm terms below only once
\[
\xi(\hat{ s}),
\quad
\xi(\mathcal{M}),
\quad
u_{\min}
\]
\vspace{0.2em}
\STATE Initialize $k \leftarrow 1$, \quad $\texttt{flag} \leftarrow 1$.
\vspace{0.2em}
\FOR{each scenario $s_k\in\mathcal{K}$}
    \STATE Compute the scenario-dependent term
    \small{
    \[
    \xi(s_{const}-\hat{s} - M|\hat{u}|)
    \]}
    \STATE Evaluate solvability conditions (\ref{eq:cond_a}) and (\ref{eq:cond_b}) for scenario $k$.
    \IF{conditions (\ref{eq:cond_a}) and (\ref{eq:cond_b}) are satisfied}
        \STATE Compute admissible radius $\rho^{(k)}$ using (\ref{SMDomain}).
        \STATE Mark scenario $k$ as solvable.
        \STATE Update $k \leftarrow k+1$.
    \ELSE
        \STATE Request full power flow solution.
        \IF{power flow iterations converge}
            \STATE Mark scenario $k$ as solvable.
            \STATE Update $k \leftarrow k+1$.
        \ELSE
            \STATE Set solvability flag $\texttt{flag} \leftarrow 0$.
            \STATE \textbf{Return} $\texttt{flag}$.
        \ENDIF
    \ENDIF
\ENDFOR
\vspace{0.2em}
\STATE Return $\texttt{flag}$.
\end{algorithmic}}
\end{algorithm}


\section{Numerical Illustration}

This section begins with a scenario featuring  a low-voltage distribution grid under high loading conditions with Volt-Var support activated. The objective is to demonstrate that during high loading conditions, excessive reactive power generation, when used to manage the voltage profile, can lead to voltage breakdown in the grid. The conditions proposed in  (\ref{eq:cond_a}) and (\ref{eq:cond_b}) are evaluated to assess power flow solvability.

In addition, the proposed conditions are evaluated on the IEEE benchmarks. The results are presented for the IEEE 34 Bus and IEEE-123 Bus feeders \cite{Kerst1}, where all power lines are adjusted to be of the same type but of different lengths, and the model parameters are based on typical values for medium voltage cables according to \cite{Kerst2}. The numerical results guarantee the existence of a solution when new loads are anticipated to be added, and voltage-dependent reactive power support via Volt-Var control is available. This shows the effectiveness of the proposed conditions for real-time decision-making, enabling efficient management of incremental grid loading under Volt-Var support provided by IBRs, without repeatedly solving power flow equations.

\subsection{5-Bus Low Voltage feeder}

\begin{figure}[htbp]
\centering
\scriptsize
\includegraphics[width=0.4\textwidth]{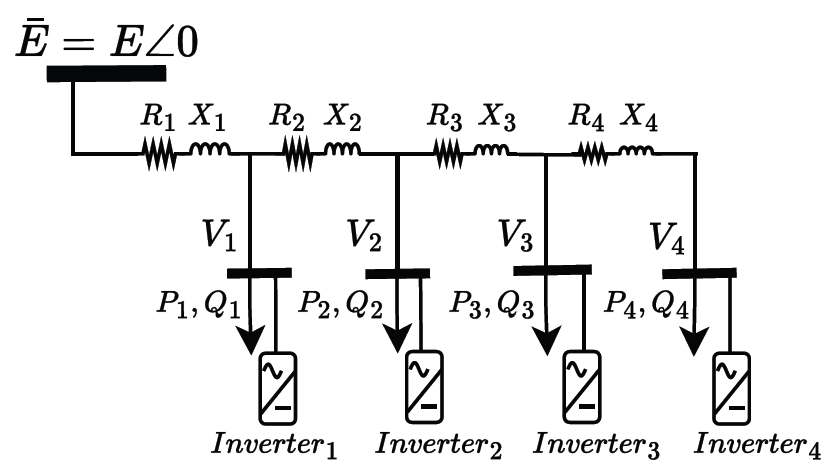}
\caption{5-Bus low-voltage feeder with Volt-Var based reactive power support.}
\label{fig:5_Bus_System}

\end{figure}

In this section, a case scenario demonstrating voltage breakdown and power flow insolvability is presented. Simulink \cite{Simulink} is used to simulate voltage collapse triggered by injection of reactive power based on Volt-Var control in a low voltage distribution feeder, as illustrated in Fig.~\ref{fig:5_Bus_System}. This setup reflects typical suburban networks. The system parameters are listed in TABLE \ref{tab: 5 Bus System Parameters}, while the inverter's setting to provide Volt-Var based reactive support is provided in TABLE \ref{tab:VoltvarSetting}. 
It is pertinent to mention that the inverters are modeled as controlled current sources. This representation ignores the dynamics of the inverter control loops, placing emphasis on the voltage support performance and load-ability limits of the connecting network.
The notation used in the table, for example $- 20$ $kVAR$ $@0.85$ $p.u.$, refers to the Volt-Var settings of the inverter defined in Fig.~\ref{fig:voltvargen}, where $Qa =-20$ $kVAR$, $Va = 0.85$ $p.u.$, $Qd = 20$ $kVAR$, $Vd = 1.15$ $p.u.$  The loading and reactive power support for each bus is provided in TABLE \ref{5 Bus System}, where positive values indicate consumption and negative values represent the generation of active/reactive power.

\begin{table}[h!]
\centering
\caption{5-Bus System Parameters}
\label{tab: 5 Bus System Parameters}
\begin{tabular}{|c|c|c|c|c|}
\hline
$E$ & 240 Volts (RMS)\\
\hline
$R_1, R_2, R_3, R_4$  & $0.07 \Omega$ \\
\hline
$X_1, X_2, X_3, X_4$  & $0.11 \Omega$ \\
\hline
\end{tabular}
\end{table}

\begin{table}[h!]
\centering
\scriptsize
\caption{Volt-Var setting for highly loaded 5-Bus system}
\label{tab:VoltvarSetting}
\begin{tabular}{|c|c|}
\hline
Bus & $Q_{\text{Volt-Var}}$ \\
\hline
1 & -20 kVAR @0.85 p.u.  \\
\hline
2 & -20 kVAR @0.85 p.u.  \\
\hline
3 & -30 kVAR @0.85 p.u. \\
\hline
4 & -35 kVAR @0.87 p.u.  \\
\hline
\end{tabular}
\end{table}

\begin{table}[h!]
\centering
\scriptsize
\caption{Key parameters for highly loaded 5 Bus System}
\label{5 Bus System}
\begin{tabular}{|c|c|c|c|c|}
\hline
Bus & P (kW) & $Q_{\text{Volt-Var}}$ (kVAR) & $|u|$(p.u.) \\
\hline
1 & 25 & -20 & 0.842 \\
\hline
2 & 25 & -20 & 0.802  \\
\hline
3 & 25 & -30 & 0.837 \\
\hline
4 & 24 & -33.4 & 0.872  \\
\hline
\end{tabular}
\end{table}

To enhance the voltage profile, it is assumed that Volt-Var settings of inverter $4$ are further adjusted by its operator. The updated settings are listed in TABLE \ref{tab:increased loading custom 5 bus}. 

\begin{figure}[htbp]
\centering
\scriptsize
\includegraphics[width=0.4\textwidth]{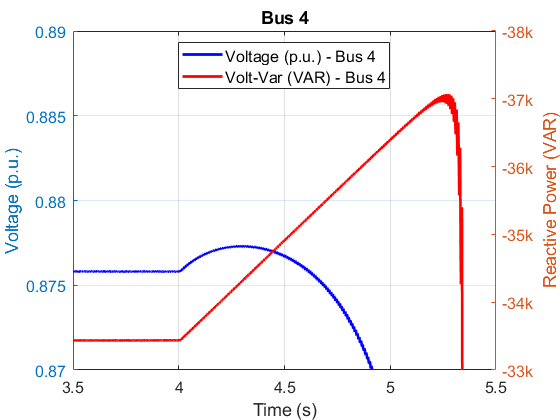}
\caption{Reactive power generation via the Volt-Var function adjusted at t=4 seconds, initially increasing voltage before a sudden drop, leading to voltage breakdown. The slew rate of 3kVAR/sec was used for reactive power injection.}
\label{fig:voltagecollapse}

\end{figure}

\begin{table}[h!]
\centering
\scriptsize
\caption{Adjusted Volt-Var Setting for highly loaded 5 Bus system}
\label{tab:increased loading custom 5 bus}
\begin{tabular}{|c|c|c|c|c|}
\hline
Bus & $\Delta Q_{\text{Volt-Var}}$  \\
\hline
4 & -38 kVAR @0.87 p.u.\\
\hline
\end{tabular}
\end{table} 
The simulation results in Figure \ref{fig:voltagecollapse}  illustrate the voltage and reactive power trajectories in response to the changes in inverter 4 settings. While the voltage at Bus 4 initially rises, the figure shows that a voltage collapse eventually occurs.

The proposed solvability criteria are applied to assess the system's behavior under increased reactive power injection. The computed values for this scenario are presented in TABLE \ref{tab:Computed Parameters - insolvable 5 Bus}, which indicates that condition \ref{eq:cond_a} is not satisfied. Although the proposed conditions are sufficient (not necessary), their violation does not imply insolvability; rather, it flags the need for a full power flow solution. This highlights the core benefit of the approach: When satisfied, the conditions enable fast screening of operating points without iterative computations. Even when violated, they provide valuable insight into the proximity to solvability boundaries, offering early warnings and guiding operational decisions.

\begin{table}[h!]
\centering
\scriptsize
\caption{Computed Parameters - Highly Loaded 5-Bus System}
\label{tab:Computed Parameters - insolvable 5 Bus}
\begin{tabular}{|c|c|c|c|}
\hline
$u_{min}$ & ${\xi(\mathcal{M})} $ & $\xi(\hat{s})$  & $\xi(s_{\text{const}} - \hat{s} - M|\hat{u}|)$  \\
\hline
0.8019 & 2.6466 & 0.8624 & 0.3371 \\
\hline
\multicolumn{1}{|c|}{L.H.S of (\ref{eq:cond_a})} & \multicolumn{3}{c|}{-2.9201} \\
\hline
\multicolumn{1}{|c|}{L.H.S of (\ref{eq:cond_b})} & \multicolumn{3}{c|}{7.1789} \\
\hline
\end{tabular}
\end{table}

For comparison, the same system is simulated under moderate loading conditions. The initial inverter settings for providing Volt-Var support, along with the power values at each bus, are presented in TABLE \ref{tab:VoltvarSetting-ModerateInit} and TABLE \ref{5 Bus System-ModerateInit}, respectively.
To enhance the voltage profile, Volt-Var control is adjusted at inverter~$4$, which has further reactive power capacity available. The updated settings are listed in TABLE \ref{tab:increased loading custom 5 bus-Moderate}. 

The results in TABLE \ref{tab:Computed Parameters - solvable 5 Bus} show that, unlike the previous case, the system satisfies both conditions (\ref{eq:cond_a}) and  (\ref{eq:cond_b}), indicating that the power flow remains solvable after adjusting the Volt-Var function.  The power values along with the voltage profile are presented in TABLE \ref{5 Bus System-ModerateAfter}.

\begin{table}[h!]
\centering
\scriptsize
\caption{Volt-Var setting for moderately loaded 5-Bus system}
\label{tab:VoltvarSetting-ModerateInit}
\begin{tabular}{|c|c|}
\hline
Bus & $Q_{\text{Volt-Var}}$ \\
\hline
1 & -5 kVAR @0.85 p.u.  \\
\hline
2 & -5 kVAR @0.85 p.u.  \\
\hline
3 & -5 kVAR @0.85 p.u. \\
\hline
4 & -1.3 kVAR @0.87 p.u.  \\
\hline
\end{tabular}
\end{table}

\begin{table}[h!]
\centering
\scriptsize
\caption{Key parameters for moderately loaded 5-Bus System }
\label{5 Bus System-ModerateInit}
\begin{tabular}{|c|c|c|c|c|}
\hline
Bus & P (kW) & $Q_{\text{Volt-Var}}$ (kVAR) & $|u|$(p.u.) \\
\hline
1 & 10 & -1.59 & 0.952 \\
\hline
2 & 10 & -2.63 & 0.921  \\
\hline
3 & 10 & -3.29 & 0.901 \\
\hline
4 & 10 & -1.10 & 0.889  \\
\hline
\end{tabular}
\end{table}

\begin{table}[h!]
\centering
\scriptsize
\caption{Adjusted Volt-Var Setting for moderately loaded 5-Bus system}
\label{tab:increased loading custom 5 bus-Moderate}
\begin{tabular}{|c|c|c|c|c|}
\hline
Bus & $\Delta Q_{\text{Volt-Var}}$  \\
\hline
4 & -3.9 kVAR @0.87 p.u.\\
\hline
\end{tabular}
\end{table}

\begin{table}[h!]
\centering
\scriptsize
\caption{Computed Parameters - Moderately Loaded 5-Bus System}
\label{tab:Computed Parameters - solvable 5 Bus}
\color{darkgreen}
\begin{tabular}{|c|c|c|c|}
\hline
$u_{min}$ & ${\xi(\mathcal{M})} $ & $\xi(\hat{s})$  & $\xi(s_{\text{const}} - \hat{s} - M|\hat{u}|)$  \\
\hline
0.8898 & 0.2716 & 0.2323 & 0.0299 \\
\hline
\multicolumn{1}{|c|}{{L.H.S of (\ref{eq:cond_a})}} & \multicolumn{3}{c|}{0.3571} \\
\hline
\multicolumn{1}{|c|}{{L.H.S of (\ref{eq:cond_b})}} & \multicolumn{3}{c|}{0.0077} \\
\hline
\multicolumn{1}{|c|}{{$\rho$ in (\ref{SMDomain})}} & \multicolumn{3}{c|}{0.1347} \\
\hline
\end{tabular}
\end{table}

\begin{table}[h!]
\centering
\scriptsize
\caption{Improved voltage profile for moderately loaded 5-Bus System with Volt/VAR control}
\label{5 Bus System-ModerateAfter}
\begin{tabular}{|c|c|c|c|c|}
\hline
Bus & P (kW) & $Q_{\text{Volt-Var}}$ (kVAR) & $|u|$(p.u.) \\
\hline
1 & 10 & -1.51 & 0.955 \\
\hline
2 & 10 & -2.46 & 0.926  \\
\hline
3 & 10 & -3 & 0.910 \\
\hline
4 & 10 & -2.93 & 0.902  \\
\hline
\end{tabular}
\end{table}
To this end, for the provision of voltage regulation service with Volt-Var function and adjusting the inverter settings, the proposed criteria can be employed to ensure the solvability of the power flow.

\subsection{37-Bus feeder}
\begin{figure}[htbp]
\centering
\includegraphics[width=0.4\textwidth]{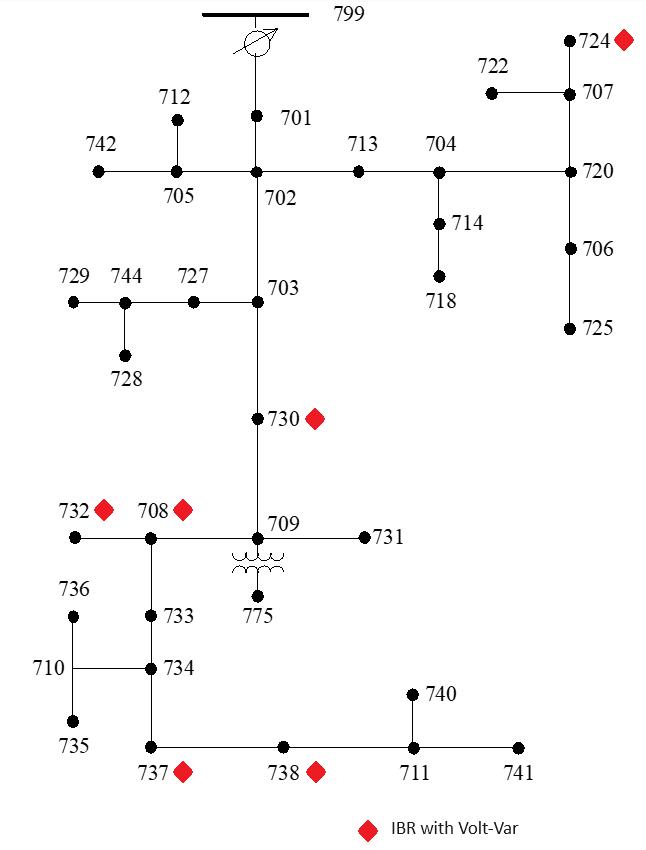}
\caption{IEEE 37 Bus feeder with location of IBRs programmed with Volt-Var.}
\label{fig:IEEE37Bus}
\end{figure}

This case focuses on assessing solvability in a low-voltage distribution network under stressed loading conditions. The structure of the IEEE 37-Bus feeder is shown in Fig.~\ref{fig:IEEE37Bus}. The base for power is $2.5$ MVA and for voltage is  $4.8/\sqrt(3)= 2.77$ kV. Buses are numbered in ascending numeric order. A single phase equivalent of the original distribution network is used for analysis \cite{Singlephase}. In constructing this model, each bus is loaded with 70\% of the maximum phase power observed in the corresponding three-phase case. The key parameters of buses prone to voltage deviation are shown in TABLE \ref{tab:key_parameters37}. It can be observed that multiple buses are at voltages lower than 0.95~p.u. At this point of operation, any further addition of load will further deteriorate the voltage profile. Providing reactive power support through Volt-Var control becomes particularly relevant in the case of further load increment, as illustrated in TABLE~\ref{tab:increased_loading37}.

\textcolor{darkgreen}{
The proposed solvability certificates are verified, as shown in TABLE \ref{tab:condition check_37}, to support operational decision-making  under Volt-Var control while guaranteeing the existence of a feasible power flow solution.
The effect of increased loading level on the voltage profiles of the most critical buses is summarized in TABLE \ref{tab:voltage_profiles_37}. The table compares bus voltage magnitudes obtained without Volt-Var support, denoted by $\tilde{u}$, and with active Volt-Var control, denoted by $u_{\text{VV}}$, along with the corresponding voltage-dependent reactive power injections, denoted by $Q_{\text{Volt-Var}}$.  The proposed solvability certificate, therefore, serves as a decision-making tool, certifying whether voltage-dependent control actions can be safely deployed without relying on iterative power-flow solvers to converge, thereby enabling real‑time decision‑making.
}

\begin{table}[htbp]
\centering
\scriptsize
\caption{Key Parameters for selected nodes of 37-Bus Feeder}
\label{tab:key_parameters37}
\begin{tabular}{|c|c|c|c|}
\hline
Bus & $P$ (kW)& $Q$ (kVAR)& $|u|$ (p.u.) \\
\hline
708 & 0 & 0 & 0.9645 \\
\hline
710 & 0 & 0 & 0.9517 \\
\hline
711 & 0 & 0 & 0.9465 \\
\hline
724 & 29.4 & 14.7 & 0.9747 \\
\hline
730 & 59.5 & 28 & 0.9719 \\
\hline
732 & 29.4 & 14.7 & 0.964 \\
\hline
733 & 59.5 & 28 & 0.9604 \\
\hline
734 & 29.4 & 14.7 & 0.9542 \\
\hline
735 & 59.5 & 28 & 0.9511 \\
\hline
736 & 29.4 & 14.7 & 0.949 \\
\hline
737 & 98 & 49 & 0.9494 \\
\hline
738 & 88.2 & 43.4 & 0.9475 \\
\hline
740 & 59.5 & 28 & 0.9459 \\
\hline
741 & 29.4 & 14.7 & 0.9462 \\
\hline
\end{tabular}
\end{table}

\begin{table}[h!]
\centering
\scriptsize
\caption{Increased Active Power loading and  Volt-Var Settings - 37-Bus}
\label{tab:increased_loading37}
\begin{tabular}{|c|c|c|}
\hline
Bus & $\Delta P$ (KW)& $\Delta Q_{\text{Volt-Var}}$\\
\hline
708 & 40 & -145kVar @ 0.90 p.u. \\
\hline
710 & 0 & 0 \\
\hline
711 & 0 & 0 \\
\hline
724 & 50 & -145 kVAR @ 0.90 p.u. \\
\hline
730 & 0 & -145 kVAR @ 0.90 p.u. \\
\hline
732 & 30 & -145 kVAR @ 0.90 p.u. \\
\hline
733 & 0 & 0 \\
\hline
734 & 40 & 0 \\
\hline
735 & 0 & 0 \\
\hline
736 & 0 & 0 \\
\hline
737 & 30 & -112.5 kVAR @ 0.90 p.u. \\
\hline
738 & 0 & -112.5 kVAR @ 0.90 p.u. \\
\hline
740 & 0 & 0 \\
\hline
741 & 0 & 0 \\
\hline
\end{tabular}
\end{table}

\begin{table}[h!]
\centering
\scriptsize
\caption{Computed Parameters -37-Bus}
\label{tab:condition check_37}
\color{darkgreen}
\begin{tabular}{|c|c|c|c|}
\hline
$u_{min}$ & ${\xi(\mathcal{M})} $ & $\xi(\hat{s})$  &$\xi(s_{\text{const}} - \hat{s} - M|\hat{u}|)$  \\
\hline
 0.9193 & 0.5116 & 0.0777 & 0.0254 \\
\hline
\multicolumn{1}{|c|}{L.H.S of (\ref{eq:cond_a})} & \multicolumn{3}{c|}{0.3231} \\
\hline
\multicolumn{1}{|c|}{L.H.S of (\ref{eq:cond_b})} & \multicolumn{3}{c|}{0.0027} \\
\hline
\multicolumn{1}{|c|}{{$\rho$ in (\ref{SMDomain})}} & \multicolumn{3}{c|}{0.1356} \\
\hline
\end{tabular}
\end{table}

\begin{table}[hbt!]
\centering
\scriptsize
\caption{Voltage Profiles and Reactive Power Support – 37-Bus}
\label{tab:voltage_profiles_37}
\begin{tabular}{|c|c|c|c|c|c|}
\hline
Bus & $P$ (KW)& $Q$ (KVar)& $|\tilde{u}|$ (p.u.) & $Q_{\text{Volt-Var}}$ (KVAR)& $|u_{\text{VV}}|$ (p.u.) \\
\hline
708 & 0+40 & 0 & 0.9568 & -46.4771 & 0.9679 \\
\hline
710 & 0 & 0 & 0.9429 & 0 & 0.9554 \\
\hline
711 & 0 & 0 & 0.9374 & 0 & 0.9511 \\
\hline
724 & 29.4+40 & 14.7 & 0.9672 & -37.4289 & 0.9742 \\
\hline
730 & 59.5 & 28 & 0.9656 & -35.5749 & 0.9755 \\
\hline
732 & 29.4+30 & 14.7 & 0.9559 & -47.3971 & 0.9673 \\
\hline
733 & 59.5 & 28 & 0.9523 & 0 & 0.964 \\
\hline
734 & 29.4+30 & 14.7 & 0.9455 & 0 & 0.9579 \\
\hline
735 & 59.5 & 28 & 0.9424 & 0 & 0.9548 \\
\hline
736 & 29.4 & 14.7 & 0.9403 & 0 & 0.9527 \\
\hline
737 & 98+20 & 49 & 0.9404 & -52.117 & 0.9537 \\
\hline
738 & 88.2 & 43.4 & 0.9384 & -53.9858 & 0.952 \\
\hline
740 & 59.5 & 28 & 0.9369 & 0 & 0.9505 \\
\hline
741 & 29.4 & 14.7 & 0.9371 & 0 & 0.9507 \\
\hline
\end{tabular}
\end{table}

\subsection{123-Bus feeder}
\begin{figure}[t]
\centering
\includegraphics[width=0.5\textwidth]{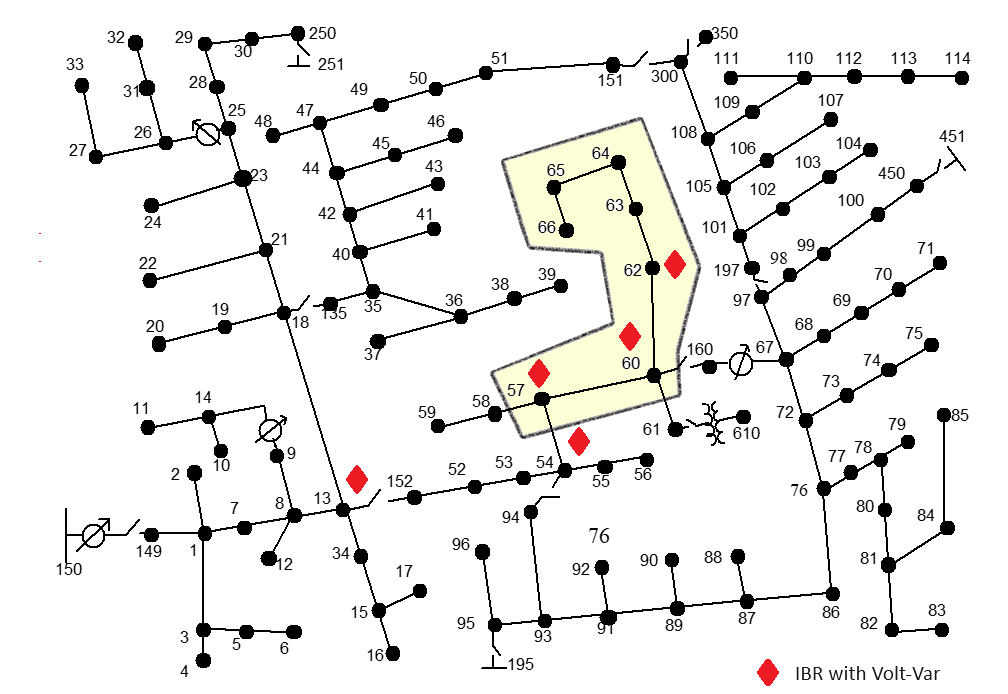}
\caption{IEEE 123-Bus feeder with the highlighted zone most prone to voltage violation and location of IBRs programmed with Volt/VAR.}
\label{fig:IEEE123Bus}
\end{figure}

This use case focuses on assessing the solvability of power flow equations in the IEEE 123-Bus distribution feeder. The base for power is $5$ MVA, and for voltage is $4.16/\sqrt(3)= 2.4$ kV. 
 Similarly to the previous case, the equivalent single-phase model of the grid is loaded with 70\% of the maximum phase power defined at each bus in the original test case. The voltage profiles of the most critical buses are shown in TABLE \ref{tab:key_parameters123}. In this test case, inverters are integrated to maintain all bus voltages above 0.95 p.u., even under a 5\% increase in load at critical buses. 
 The network layout shown in Fig.~\ref{fig:IEEE123Bus} highlights the buses most affected by under voltage conditions, which are encircled, along with the locations of Volt-Var support capable IBRs at five designated points. 

At this point, the operator decides to turn on the Volt-Var support at these five designated points. The solvability of power flow with activated Volt-Var functions with the setting of -180 kVAR @ 0.90 p.u. and an increase of active power load is of particular interest. 

\textcolor{darkgreen}{
The proposed solvability certificates are utilized to make operational decisions for activating Volt–Var control under increased loading conditions. The parameters calculated for the corresponding solvability analysis are reported in TABLE \ref{tab:condition check_123}, which guarantees the existence of a feasible power flow solution.
The resulting voltage profiles at the most critical buses, as well as at buses hosting inverter-based resources under full loading conditions, are summarized in TABLE \ref{tab:voltage_profiles_WithVoltvar123}. The table compares bus voltage magnitudes obtained without Volt–Var support, denoted by $\tilde{u}$, and with active Volt–Var control, denoted by $u_{\text{VV}}$, together with the corresponding voltage-dependent reactive power injections provided by the inverters, denoted by $Q_{\text{Volt-Var}}$. The proposed conditions, therefore, serve as an active decision-making tool for enabling or tuning voltage regulation services.
}

\begin{table}[t]
\centering
\scriptsize
\caption{Key Parameters for selected nodes of 123-Bus Feeder}
\label{tab:key_parameters123}
\begin{tabular}{|c|c|c|c|}
\hline
Bus & $P$ (KW)& $Q$ (KVAR)& $|u|$ (p.u.) \\
\hline
66 & 52.5 & 24.5 & 0.937333 \\
\hline
65 & 49 & 35 & 0.938075 \\
\hline
64 & 52.5 & 24.5 & 0.940975 \\
\hline
63 & 28 & 14 & 0.944159 \\
\hline
62 & 28 & 14 & 0.945965 \\
\hline
160 & 0 & 0 & 0.948847 \\
\hline
60 & 14 & 7 & 0.948995 \\
\hline
61 & 0 & 0 & 0.948995 \\
\hline
114 & 14 & 7 & 0.960664 \\
\hline
113 & 28 & 14 & 0.960976 \\
\hline
\end{tabular}
\end{table}

\begin{table}[hbt!]
\centering
\scriptsize
\caption{Computed Parameters - 123-Bus}
\label{tab:condition check_123}
\color{darkgreen}
\begin{tabular}{|c|c|c|c|}
\hline
$u_{min}$ & ${\xi(\mathcal{M})} $ & $\xi(\hat{s})$  &$\xi(s_{\text{const}} - \hat{s} - M|\hat{u}|)$  \\
\hline
 0.8825 & 0.4480 & 0.1260 & 0.0196\\
\hline
\multicolumn{1}{|c|}{L.H.S of (\ref{eq:cond_a})} & \multicolumn{3}{c|}{0.2917} \\
\hline
\multicolumn{1}{|c|}{L.H.S of (\ref{eq:cond_b})} & \multicolumn{3}{c|}{0.0067} \\
\hline
\multicolumn{1}{|c|}{{$\rho$ in (\ref{SMDomain})}} & \multicolumn{3}{c|}{0.1049} \\
\hline
\end{tabular}
\end{table}

\begin{table}[hbt!]
\centering
\scriptsize
\caption{Voltage Profiles and Reactive Power Support - 123-Bus}
\label{tab:voltage_profiles_WithVoltvar123}
\begin{tabular}{|c|c|c|c|c|c|}
\hline
Bus & $P$ (KW)& $Q$ (KVAR)& $\tilde{u}$ (p.u.) & $Q_{\text{Volt-Var}}$(KVAR)& $u_{\text{VV}}$ (p.u.) \\
\hline
66 & 56.25 & 24.5 & 0.935985 & 0 & 0.950342 \\
\hline
65 & 52.5 & 35 & 0.936771 & 0 & 0.951116 \\
\hline
64 & 56.25 & 24.5 & 0.939784 & 0 & 0.954099 \\
\hline
63 & 30 & 14 & 0.943108 & 0 & 0.957385 \\
\hline
62 & 30 & 14 & 0.944997 & -73.3477 & 0.959251 \\
\hline
160 & 0 & 0 & 0.948015 & 0 & 0.961935 \\
\hline
60 & 15 & 7 & 0.948163 & -68.2577 & 0.962079 \\
\hline
61 & 0 & 0 & 0.948163 & 0 & 0.962079 \\
\hline
114 & 15 & 7 & 0.959534 & 0 & 0.974202 \\
\hline
113 & 30 & 14 & 0.959860 & 0 & 0.974524 \\
\hline
112 & 14 & 7 & 0.961928 & 0 & 0.976567 \\
\hline
57 & 0 & 0 & 0.964412 & -43.6894 & 0.975728 \\
\hline
54 & 0 & 0 & 0.972312 & -32.2065 & 0.982108 \\
\hline
13 & 0 & 0 & 0.990269 & -6.4634 & 0.996409 \\
\hline
\end{tabular}
\end{table}

\section{Conclusion}

This paper presents a novel extension to power flow solvability analysis that incorporates Volt-Var based reactive power support from IBRs. The proposed framework assists in real-time decision making under varying operating conditions, which are often computationally demanding, especially when evaluating the feasibility of new operating points. The approach is validated using positive-sequence equivalent models of the IEEE 37-bus and 123-bus test feeders. Results demonstrate that the proposed condition accurately predicts the existence of power flow solutions under stressed operating conditions, highlighting its potential for real-time voltage regulation applications utilizing a network of IBRs.
\textcolor{darkgreen}{
Future work will focus on reducing conservativeness through tighter norm bounds and refined normalization, while extending the proposed solvability framework to explicitly account for unbalanced multiphase distribution networks.
}

\newpage

\end{document}